# Quantifying Public Response to COVID-19 Events: Introducing the Community Sentiment and Engagement Index

Nirmalya Thakur
Department of Electrical Engineering and Computer Science
South Dakota School of Mines and Technology
Rapid City, SD 57701, USA
nirmalya.thakur@sdsmt.edu

Kesha A. Patel
Department of Mathematics
Emory University
Atlanta, GA 30322, USA
kesha.patel@emory.edu

Audrey Poon
Department of Computer Science
Emory University
Atlanta, GA 30322, USA
audrey.poon@emory.edu

Shuqi Cui
Department of Computer Science
Emory University
Atlanta, GA 30322, USA
nicole.cui@emory.edu

Nazif Azizi
Department of Computer Science
Emory University
Atlanta, GA 30322, USA
mohammad.nazif.azizi@emory.edu

Rishika Shah
Department of Computer Science
Emory University
Atlanta, GA 30322, USA
rishika.shah@emory.edu

Riyan Shah
Department of Computer Science
Emory University
Atlanta, GA 30322, USA
riyan.shah@emory.edu

*Abstract*— **This study introduces the Community Sentiment and Engagement Index (CSEI), developed to capture nuanced public sentiment and engagement variations on social media, particularly in response to major events related to COVID-19. Constructed with diverse sentiment indicators, CSEI integrates features like engagement, daily post count, compound sentiment, fine-grain sentiments (fear, surprise, joy, sadness, anger, disgust, and neutral), readability, offensiveness, and domain diversity. Each component is systematically weighted through a multi-step Principal Component Analysis (PCA)-based framework, prioritizing features according to their variance contributions across temporal sentiment shifts. This approach dynamically adjusts component importance, enabling CSEI to precisely capture high-sensitivity shifts in public sentiment. CSEI's development showed statistically significant correlations with its constituent features, underscoring internal consistency and sensitivity to specific sentiment dimensions. CSEI's responsiveness was validated using a dataset of 4,510,178 Reddit posts about COVID-19. The analysis focused on 15 major events, including the WHO's declaration of COVID-19 as a pandemic, the first reported cases of COVID-19 across different countries, national lockdowns, vaccine developments, and crucial public health measures. Cumulative changes in CSEI revealed prominent peaks and valleys aligned with these events, indicating significant patterns in public sentiment across different phases of the pandemic. Pearson correlation analysis further confirmed a statistically significant relationship between CSEI daily fluctuations and these events (p = 0.0428), highlighting CSEI's capacity to infer and interpret the shifts in public sentiment and engagement in response to major events related to COVID-19.**

I. INTRODUCTION

In December 2019, an outbreak of coronavirus disease 2019 (COVID-19) due to severe acute respiratory syndrome coronavirus 2 (SARS-CoV-2) began in China. [1]. After the initial outbreak, COVID-19 soon spread to different parts of the world, and on March 11, 2020, the World Health Organization (WHO) declared COVID-19 an emergency [2]. As of October 27, 2024, there have been a total of 776,754,317 cases and 7,073,466 deaths due to COVID-19 [3]. The COVID-19 pandemic has posed not only a severe public health crisis but also an extensive psychological and social challenge worldwide, affecting nearly every aspect of life. Beyond the devastating physical health implications, the pandemic disrupted social norms, created economic uncertainty, and intensified psychological stress across communities [4,5]. Faced with isolation measures, travel restrictions, and heightened anxieties about personal and community health, individuals increasingly turned to social media platforms as spaces for emotional expression, information-sharing, and social connection [6,7].

Several social media platforms, such as Facebook, Twitter, Instagram, Reddit, TikTok, and YouTube, have become very popular since the beginning of COVID-19 [8,9]. Out of these social media platforms, Reddit, in particular, stands out for its unique features that allow focused discussions via subreddits. The subreddits are often categorized by topic and interest. Reddit allows users to post content to which other users can respond, vote, and discuss. Reddit users may join any number of subreddits related to their interests. From July 2023 to December 2023, Reddit consistently amassed at least 1.8 billion monthly visits [10]. As of December 2023, the top five countries in terms of number of Reddit users were United States (48.46%), United Kingdom (7.17%), Canada (6.97%), Australia (4.05%), and Germany (3.22%) [11]. In January 2024, reddit.com received approximately 4.45 billion visits from users in the United States, making it the top-ranking country in terms of online traffic to the platform. Canada followed closely behind with 376.7 million visits, while the United Kingdom secured the third position with nearly 325 million visits during the same month [12]. In January 2024, reddit.com received almost 6 billion visits worldwide through mobile devices and 1.67 billion visits through desktop devices. In addition to this, various social media platforms refer users to reddit.com. In December 2023, youtube.com was the leading referrer and accounted for over half of the referral traffic at (56.81%), X (formerly Twitter) was the second social media traffic driver at 20.72%, and the remaining social media traffic drivers to reddit.com were facebook.com, whatsapp.com, and instagram.com at 7.78%, 1.84%, and 1.8% respectively [13]. Reddit currently has 1.2 billion users on a global scale [14], and on average, a Reddit user spends 1089 seconds on the platform per session [15].

Given the extensive and varied discourse on platforms like Reddit, analyzing community sentiment and engagement has become crucial for researchers seeking to understand public morale, views, and opinions toward the pandemic [16]. There have been several works conducted in this field that analyzed the public discourse about COVID-19 on social media platforms such as Facebook [17,18], Twitter [19,20], YouTube [21,22], WhatsApp [23,24], Instagram [25,26], TikTok [27,28], WeChat [29,30], and Weibo [31,32] since the beginning of the pandemic. However, there is still limited research about investigating and analyzing the public discourse about COVID-19 on Reddit. Therefore, this work used a dataset of 4,510,178 Reddit posts about COVID-19 [33] to develop and evaluate CSEI. Metrics that reflect community sentiment and engagement offer a window into public well-being and can serve as early indicators of community needs, anxieties, and responses to governmental policies and health guidelines. Such insights are invaluable for public health researchers, policymakers, and mental health professionals aiming to make data-driven decisions to support public well-being. However, traditional sentiment analysis approaches on social media tend to focus narrowly on basic sentiment polarity, i.e., whether the content is positive, negative, or neutral, thereby offering only a limited view of the collective emotional state [34-39]. This narrow focus fails to capture the breadth of public sentiment, as it overlooks critical dimensions such as the diversity of emotional responses, the quality of discourse, and varying levels of engagement that shape and reflect community well-being. These limitations underscore the need for a comprehensive metric encompassing multiple facets of online discourse.

To address this gap, this paper introduces the Community Sentiment and Engagement Index (CSEI). CSEI is a metric specifically designed to capture the complex and multi-dimensional nature of community health in the public discourse on social media platforms such as Reddit. The CSEI integrates a broad array of features, including compound sentiment scores, daily post frequency, content readability, offensiveness, domain diversity, and fine-grained emotional components such as fear, surprise, joy, sadness, anger, disgust, and neutral. Each of these components contributes uniquely to a comprehensive measure that captures the overall state of community well-being. For instance, compound sentiment provides an aggregate measure of overall positivity or negativity, while fine-grained emotions reveal specific emotional responses. Engagement metrics such as daily post count and domain diversity capture the community's level of participation and the range of sources that shape the narrative. Content quality indicators, like readability and offensiveness, reveal the tone and accessibility of discourse, adding additional layers to the analysis. Together, these diverse dimensions form an index that not only assesses sentiment but also accounts for the civility, accessibility, and engagement of the community's online discourse.

The process of developing CSEI began with feature normalization, essential for balancing variables with differing ranges, preventing high-variance features like post frequency and domain diversity from overshadowing subtle yet critical aspects like fine-grained sentiments. Following this, Principal Component Analysis (PCA) [40] was applied to assess each feature's contribution based on variance, allowing the extraction of principal components that encapsulate the most informative dimensions of sentiment dynamics. By focusing on the first principal component, weights were derived to emphasize high-loading features, enhancing the CSEI's responsiveness to influential aspects of the discourse. The final index was created through weighted linear aggregation [41], scaling each feature by its PCA-determined statistical significance. This approach produced an index that is not only sensitive to shifts in public sentiment but also captures the complex, multi-dimensional nature of the public discourse on social media platforms, such as Reddit. The applicability of the CSEI was validated by examining its trends over time in response to 15 major events related to COVID-19. CSEI captured immediate sentiment responses, while a moving average filtered noise, highlighting sustained trends. Peaks and valleys associated with these events were identified using prominence criteria, aligning with events like lockdowns, vaccine rollouts, and global health alerts, underscoring the CSEI's sensitivity to temporal sentiment dynamics. Pearson correlation analysis [42] demonstrated a statistically significant correlation between CSEI and all these major events related to COVID-19, confirming the CSEI's effectiveness in inferring and interpreting shifts in public sentiment and engagement in response to major COVID-19 developments.

The rest of this paper is presented as follows. Section 2 reviews the recent works in this field. Section 3 presents the methodology that was followed for the development and evaluation of CSEI. Section 4 presents the results and highlights the scientific contributions of this work. It is followed by Section 5, which concludes the paper and outlines the scope for future work in this field.

II. LITERATURE REVIEW

The analysis and investigation of public discourse about COVID-19 on social media platforms such as Reddit has attracted the attention of researchers from different disciplines in the recent past. Multiple works have analyzed the patterns of sentiment, engagement, and other factors related to seeking and sharing information about COVID-19 on Reddit.

Hu et al. [43] examined public attitudes toward COVID-19 in the United States, the United Kingdom, and Australia. They utilized data from Reddit, specifically country and COVID-19-related subreddits (r/CoronavirusAustralia, r/CoronavirusDownunder, r/CoronavirusCanada, r/CanandaCoronavirus, r/coronavirus) for their work. Their dataset contained 84,229 posts and 1,094,853 comments published between February 2020 and November 2020. They used topic modeling to deduce and compare the topics of concern related to COVID-19 for each country. The results showed that COVID-19 posts declined during the study period; however, during lockdown, there was an increase in posts. Moreover, the UK and Australian subreddits contained more evidence-based policy discussions. Hale et al. [44] investigated the dialogue in the subreddit r/coronavirus. The authors analyzed 226 posts and 2260 comments published between February 2020 and May 2020. The findings of their work showed that users valued information regarding COVID-19 spread, public health information, political and economic implications of COVID-19, and the experiences of medical workers. Moreover, the researchers discovered that posts were collectively oriented and negatively valenced. Thompson et al. [45] studied 31,892 posts from the subreddit r/covidlonghaulers, published between July 24, 2020 and January 7, 2021. Their investigation revealed 16 distinct factors of words that described symptoms, diagnostic concerns, general health-related concerns, chronicity, support, identity, and anxiety. Lee et al. [46] utilized Reddit to identify 11,830 comments related to foster families to study their well-being during COVID-19. The findings of their study indicated an increase in discussions about becoming a foster parent and activities for foster children during COVID-19. Sarker et al. [47] aimed to identify self-reported COVID-19 symptoms, compare symptom distributions, and develop a collection of COVID-19 symptoms. The authors collected 42,995 posts from 4249 users from the r/covidlonghaulers subreddit. They discovered the most frequently reported long-COVID symptoms were mental health-related, fatigue, general ache or pain, brain fog, confusion, and dyspnea. Through the use of exponential random graph modeling, the work of Chipidza et al. [48] inferred that homophily and toxic posts about COVID-19 often occurred simultaneously on Reddit. Slemon et al. [49] analyzed Reddit posts published from February 2020 to December 2020 on the subreddit forum r/COVID19_support. Their work aimed to perform thematic analysis to understand the themes and main concerns with a specific focus on interpreting mental health and suicidal thought-related communications.

Basile et al. [50] investigated Reddit mega threads and used the Plutchik model of basic emotions to compare the reactions and structures of Reddit posts impacted and not impacted by COVID-19. The findings of their work showed that emotions on Reddit increased during events related to COVID-19. Wanchoo et al. [51] studied one million Reddit posts about COVID-19 published between January 6, 2019, and January 5, 2021, belonging to different subreddits regarding diet, physical activity, substance use, and smoking. Their work showed that due to COVID-19, there was a considerable variation in the subject matter of posts. More specifically, before the pandemic, common topics involved vacation, international travel, work, family, consumption of illicit substances, vaping, and alcohol. During the pandemic, common topics were quarantine, withdrawal symptoms, anxiety, attempts to quit smoking, cravings, weight loss, and physical fitness. Their work also showed that discussions about benzodiazepines, opioids, and quitting smoking peaked around March 2020. Alambo et al. [52] conducted a longitudinal topical analysis of Reddit posts about COVID-19 published between January 2020 and October 2020 on different subreddits. Their study showed that in September 2020, there was a high topical correlation between postings in r/depression and r/Coronavirus.

In [53], Jai et al. investigated a dataset of Reddit posts related to COVID-19 to explore gender differences in language use. Their work specifically focused on analyzing the lexical, topical, and emotional differences. Wu et al. [54] aimed to investigate the concerns of different communities regarding COVID-19 vaccinations by exploring Reddit posts. The findings of their work showed that most posts were overwhelmed with content about conspiracy theories, each subreddit had its unique audience (some of which were mutually exclusive with other subreddits, and thus information was often stagnant), and when users' discussions fluctuated with time and the specific subreddit - the communication strategies also reflected these changes. Veselovsky et al. [55] proposed a strategy to examine the evolution of Reddit during the pandemic. The results of their work revealed that Reddit grew throughout various communities and languages, with older users incorporating COVID-related language more frequently than new users. Moreover, they observed that new users also had radically different interests and activities. Kimiafar et al. [56] used multiple concepts of data analysis and information retrieval to understand the concerns of the global population regarding COVID-19, as expressed on Reddit. In view of the impact COVID-19 had on people's lives and the influence of social media during social distancing, Whitfield et at. [57] used concepts of named-entity recognition (NER) and Latent Dirichlet Allocation (LDA) to interpret the discourse on Reddit about COVID-19. The findings of their work showed that the most commonly debated topics on Reddit were related to testing, masks, and employment. Murray et al. [58] utilized topic modeling and sentiment analysis to analyze posts from the subreddit r/COVID19Positive. Posts published on this subreddit forum involved information about positive symptoms and personal struggles regarding the contraction of the virus. The results of their work showed that critical cases of the virus were discussed at a continuous rate on Reddit. They were also able to infer that breathing issues reached their peak around ten days after testing positive for COVID-19 by interpreting the content of these posts. Wang et at. [59] aimed to explore mental health issues such as depression and anxiety among students attending historically Black colleges and universities during the COVID-19 pandemic. Their study showed that 49% of students expressed symptoms of depression, and 52% expressed a combination of depression and anxiety.

In summary, prior works in this field have primarily relied on topic modeling, thematic analysis, and sentiment analysis to understand COVID-19-related discussions, focusing on specific emotions, mental health concerns, or social behaviors within limited datasets and often confined to particular subreddits. This narrow focus lacks a multi-dimensional measure that integrates diverse emotional tones, engagement indicators, and sentiment nuances over time to provide a comprehensive understanding of community well-being across varied contexts. Furthermore, studies analyzing the dynamics of public sentiment generally used basic sentiment polarity (such as positive, negative, and neutral), which does not capture the full spectrum of community responses to ongoing events. The CSEI addresses these gaps by offering a robust approach that combines sentiment, fine-grained emotional markers, engagement metrics, and discourse quality indicators to quantify community sentiment and engagement dynamically. Finally, the dataset that was used to develop and evaluate the effectiveness of CSEI contains

4,510,178 Reddit posts about COVID-19, and the number of posts present in this dataset is considerably higher than the number of Reddit posts studied in any prior work in this field. The step-by-step process that was followed for the development and evaluation of CSEI is described in Section 3, and the results are presented in Section 4.

### III. METHODOLOGY

The dataset that was used for this research work contains 4,510,178 distinct Reddit posts about COVID-19 [33] published on 101,862 subreddits. The top 20 subreddits in this dataset in terms of the number of posts are u_toronto_news (18.86%), autonewspaper (3.58%), coronavirus (2.41%), askreddit (1.88%), news (1.56%), covid_canada (1.26%), newsbotbot (1.24%), worldnews (0.98%), ddnews (0.92%), politics (0.68%), conspiracy (0.63%), innews (0.53%), talkativepeople (0.49%), covid19 (0.43%), stardiapostcom (0.39%), thenewsfeed (0.38%), showerthoughts (0.36%), nonewnormal (0.35%), nofeenews (0.34%), and india (0.32%). The date range of this dataset was observed to be February 11, 2020, to October 25, 2021, following the exclusion of seven posts with unusually early dates (i.e., prior to December 2019) where the word covid or one of its synonyms was present but in a completely different context. Table 1 presents a data description of this dataset.

Table 1: Data Description of the dataset

| Attribute Name | Description |
|---|---|
| type | Type of Reddit entry (e.g., 'post') |
| id | Unique identifier for each post |
| subreddit.id | Identifier for the subreddit where the post was published |
| subreddit.name | Name of the subreddit |
| subreddit.nsfw | Indicates if the subreddit is marked as NSFW (not safe for work) |
| created_utc | UTC timestamp for when the post was published |
| permalink | URL path of the post on Reddit |
| domain | Domain source of the post's content |
| url | URL of the post's linked content |
| selftext | Text of the post |
| title | Title of the post |
| score | Score of a post, computed as upvotes - downvotes |

The data preprocessing included multiple steps. First, deleted posts (marked as "[deleted]" or "[removed]" in the dataset) were removed. This was followed by the detection and removal of bot-generated content. In this dataset, any bot-generated content contained the phrase – "I am a bot". Thereafter, the next set of data preprocessing steps that were applied to the posts included the detection of English words, removal of non-alphabetic characters, URLs, hashtags, user mentions, stop words, and digits. After data preprocessing, outlier detection and removal were performed. The outlier detection and removal involved two steps. First, the Isolation Forest approach [60] was used. Configured with a contamination rate of 0.5%, Isolation Forest identified observations that deviated significantly from the norm, capturing general outliers, which were then deleted. Second, the Principal Component Analysis (PCA) was used. Specific outliers were identified based on principal component scores: observations with a score below 25 for the first principal component (PC1) and a score of at least 7.5 for the second (PC2) were flagged as additional anomalies and removed.

After the completion of data preprocessing and outlier removal, fine-grain sentiment analysis, offensive score computation, readability analysis, and sentiment analysis were performed on each post to generate the additional columns required for the development of the CSEI. Fine-Grain sentiment analysis was performed using the "j-hartmann/emotion-english-distilroberta-base" model, which is based on DistilRoBERTa [61]. DistilRoBERTa functions by encoding text into dense vector representations using a transformer architecture, distinguishing linguistic patterns through attention mechanisms. This model generated probability distributions for seven emotions (fear, surprise, joy, sadness, anger, disgust, and neutral) for each post, utilizing the highest probability score to apply a sentiment label. The offensive score was calculated utilizing the "cardiffnlp/twitter-roberta-base-offensive" model [62], a variant of RoBERTa, specifically fine-tuned for offensive language analysis. This model tokenizes input text, processes it via several transformer layers, and utilizes a softmax function [63] on the logits of each class to produce probabilities. The results indicated the probability of each post containing offensive language, crucial for assessing discourse quality with a specific focus on the degree of civility. The Flesch Reading Ease test [64], a widely accepted metric for assessing text difficulty, was utilized in readability analysis using the readability library [65]. This method assesses readability by evaluating average sentence length and syllable count per word, where higher scores signify a higher degree of comprehension. Finally, the compound sentiment score was computed utilizing VADER (Valence Aware Dictionary and sEntiment Reasoner) [66], which is specifically designed for the analysis of social media vocabulary and has been used in several prior works in this field [67-70]. VADER functions via a vocabulary and rule-based methodology, attributing sentiment intensities to words and employing heuristics for punctuation, capitalization, and degree modifiers. The results from these four models were added as separate attributes in the dataset.

Thereafter, the step-by-step process to develop the CSEI was started. The CSEI development process comprised three fundamental steps: feature normalization, Principal Component Analysis (PCA)-based weighting, and weighted linear aggregation. Feature normalization was essential to the process, enabling features with different magnitudes and units to contribute uniformly to the final index. Min-max normalization was used to standardize each feature $X(i,t)$ at time $t$ onto a uniform scale, rescaling it to the interval $[0,1]$ shown in Equation 1, where $\min(X(i))$ and $\max(X(i))$ represented the minimum and maximum values of $X(i)$ across the dataset.

$$X_{norm} = \frac{X(i,t) - \min(X(i))}{\max(X(i)) - \min(X(i))} \qquad (1)$$

The PCA-based weighting was crucial in the development of the CSEI, as it facilitated the allocation of weights to features according to their contribution to overall variance. PCA, an eigenvalue decomposition technique [71], was used to identify the principle components, each signifying a direction of maximal variation within the dataset. This method efficiently diminished dimensionality, preserving solely the most significant components. In this context, the first principal component (PC1) was notably relevant, as it encapsulated the direction of maximal variation, enabling each feature's weight to represent its contribution to the sentiment and engagement dynamics represented by the CSEI. The theoretical basis of PCA as an eigenvalue decomposition method established a systematic approach to weighting, wherein a specific feature X(i) was allocated a weight w(i), as demonstrated in Equation 2, where l(i) represented the loading of feature X(i) in PC1.

$$w(i) = \frac{|l(i)|}{\sum_{j=1}^{n} |l(j)|} \quad (2)$$

The PCA-based weighting in the development of the CSEI allocated more weights to features that captured significant variance, thereby emphasizing their relevance in characterizing the sentiment and engagement patterns present in the data. The loading of each characteristic in PC1 directly influenced its allocated weight, creating a systematic and variance-informed basis for developing the CSEI. This PCA method provided dimensional reduction and weighted aggregation, allowing the CSEI to be interpretable and responsive to the dataset's principal sources of variance. The CSEI was calculated using weighted linear aggregation, which integrated the PCA-derived weights and normalized features into a single index. At each time t, the CSEI represented as CSEI(t), was computed as the aggregate of the normalized values of each feature X(i,t)norm, weighted by w(i), as illustrated in Equation 3.

$$CSEI(t) = \sum_{i=1}^{n} w(i) \cdot X(i, t)_{norm} \quad (3)$$

Metrics related to community engagement, including daily post count and domain diversity, were essential elements of the CSEI, reflecting the extent of public participation and the range of topics present in online discussions. The engagement metrics are shown in Equation 4.

$$\text{Engagement Component} = f(\text{Post Count, Domain Diversity}) \quad (4)$$

Furthermore, qualitative metrics, such as readability and offensive content, played a role in the CSEI by reflecting dimensions of information clarity and civility. The qualitative components were compiled, as shown in Equation 5. Equation 6 shows the generic form of CSEI. The specific formula of CSEI and the weights associated with each of these features are presented in Section 4.

$$\text{Quality Component} = f(\text{Readability, Offensive Content}) \quad (5)$$

$$CSEI(t) = \alpha.\text{Sentiment Component} + \beta.\text{Engagement Component} + \gamma.\text{Quality Component} \quad (6)$$

To analyze how the CSEI responded to COVID-19-related events, the daily changes in CSEI were studied. Let $CSEI_t$ denote the time series of CSEI values over t days. The daily change, $\Delta CSEI_t$, was computed as the first-order difference, isolating the day-to-day shifts in community sentiment and engagement, as shown in Equation 7.

$$\Delta CSEI_t = CSEI_t - CSEI_{t-1} \quad (7)$$

Given that daily changes in CSEI may exhibit high variance, likely due to noise, the series was smoothed by applying a moving average over a window w [72]. The rolling mean for each day t, calculated over a 7-day window, provided a clearer view of trends by filtering out high-frequency variations, as presented in Equation 8.

$$CSEI_{smooth}(t) = \frac{1}{w} \sum_{k=t-w+1}^{t} \Delta CSEI_k \quad (8)$$

Identifying significant peaks and valleys [73] within the smoothed CSEI series was essential for understanding local maxima and minima in public sentiment and engagement. A peak at time t was defined as a point where the value of CSEIsmooth(t) was greater than neighboring values within a specified distance d. This condition for a peak is shown in Equation 9. For the peak to be significant, this condition had to hold over subsequent days within the range d, as shown in Equation 10. Furthermore, the prominence of the peak was assessed by applying a minimum threshold p to filter out inconsequential peaks, as indicated in Equation 11.

$$CSEI_{smooth}(t) > CSEI_{smooth}(t \pm d) \quad (9)$$

$$CSEI_{smooth}(t) > CSEI_{smooth}(t + k) \text{ for } k = 1, 2, \ldots, d \quad (10)$$

$$CSEI_{smooth}(t) - \min(CSEI_{smooth}(t - d : t + d)) \geq p \quad (11)$$

Valleys were identified similarly by applying analogous criteria to the negated smoothed CSEI series. A valley at time t satisfied the condition shown in Equation 12. This condition also required validation over subsequent values, as specified in Equation 13. The cumulative impact of COVID-19 events on the CSEI was captured by computing the cumulative change in CSEI over time. The cumulative change at time t, represented as $CumulativeChange_t$, was calculated by summing daily changes up to t, as shown in Equation 14.

$$CSEI_{smooth}(t) < CSEI_{smooth}(t \pm d) \quad (12)$$

$$CSEI_{smooth}(t) < CSEI_{smooth}(t + k) \text{ for } k = 1, 2, \ldots, d \quad (13)$$

$$CumulativeChange_t = \sum_{k=1}^{t} \Delta CSEI_k \quad (14)$$

To analyze the influence of specific events on cumulative sentiment and engagement shifts, an event indicator $E_t$, which took a value of 1 on event dates and 0 otherwise, was introduced, as shown in Equation 15. Finally, Pearson's p-value was used to quantify the statistical significance of the relationship between changes in CSEI and major events to infer the capacity of CSEI to detect and interpret public sentiment shifts in response to major COVID-19 developments.

$$E_t = 1 \text{ (if t is an event date)}, \quad 0 \text{ (otherwise)} \quad (15)$$

Fifteen major global events related to COVID-19 [74] were analyzed in this study. These events are stated as follows:

- February 11, 2020: WHO officially named the novel coronavirus disease as COVID-19 [75]

- March 11, 2020: WHO declared COVID-19 a pandemic [76]
- April 7, 2020: The Wuhan lockdown ended [77]
- May 5, 2020: The UK reported the highest COVID-19 death toll in Europe [78]
- June 11, 2020: COVID-19 cases began to surge across Africa [79]
- July 30, 2020: COVID-19 reached severe levels globally [80]
- September 28, 2020: Global COVID-19 deaths surpassed 1 million [81]
- November 9, 2020: Pfizer announced the efficacy of its COVID-19 vaccine [82]
- December 2, 2020: The UK became the first country to approve the Pfizer vaccine for emergency use [83]
- January 13, 2021: CDC issued a warning about potential COVID-19 surges [84]
- March 11, 2021: COVID-19 deaths globally reached 2.6 million [85]
- April 25, 2021: India experienced a severe second COVID-19 wave [86]
- June 24, 2021: South Africa faced its third COVID-19 wave [87]
- August 3, 2021: The CDC reported that over 90% of new infections were due to the Delta variant in the US [88]
- September 14, 2021: WHO recommended delaying booster shots for fully vaccinated individuals to address vaccine equity for countries with low vaccination rates [89].

A program was written in Python 3.10 as per the above methodology for developing and evaluating CSEI. The detailed results and findings derived from this approach are presented in Section 4.

## IV. RESULTS AND DISCUSSIONS

This section presents the results of this research work. First, the equation for CSEI is presented, which was developed as per the step-by-step methodology discussed in Section III. The CSEI equation, presented in Equation 16, serves as a composite measure designed to capture the multifaceted nature of public sentiment and engagement

$$CSEI = 0.1398 \cdot CompoundSentiment + 0.0057 \cdot daily\_total\_score + 0.1183 \cdot daily\_post\_count + 0.0438 \cdot readability + 0.1386 \cdot offensive + 0.1761 \cdot domain\_diversity + 0.0200 \cdot anger + 0.0120 \cdot disgust + 0.0731 \cdot fear + 0.0830 \cdot joy + 0.0622 \cdot neutral + 0.0828 \cdot sadness + 0.0447 \cdot surprise \quad (16)$$

Every element of the formula signifies a unique aspect of public sentiment or engagement, with coefficients obtained from the program that captures the impact of each factor on the overall index. For example, domain_diversity, which reflects the range of online sources that contribute to public discourse, has the highest coefficient (0.1761). This suggests its influence in expanding the reach and diversity of information related to COVID-19. A broad range of domains is associated with increased engagement and various viewpoints, indicating a higher public reaction. The offensive component, with a coefficient of 0.1386, indicates the degree of offensive content or civility in the discourse, serving as a crucial measure of the potential tension or polarization in discussions related to the pandemic. Heightened emotions and polarization are typically associated with increased offensive content, particularly evident in discussions surrounding topics like lockdown measures, vaccine mandates, and various pandemic-related interventions. Meanwhile, compound sentiment, with a coefficient of 0.1398, captures the general sentiment tone and reflects shifts in public emotions over time. Increased values in compound sentiment reflect a prevailing trend of optimism or pessimism, directly correlating with the overall emotional reaction to developments related to the pandemic. The findings of Pearson's correlation analysis of CSEI with each of its constituent features are shown in Figure 1, where the r-values are shown. Person's p-values were also computed, and the findings showed that CSEI has a statistically significant correlation with each of its constituent terms as the p-values for each of these correlations were less than 0.05.

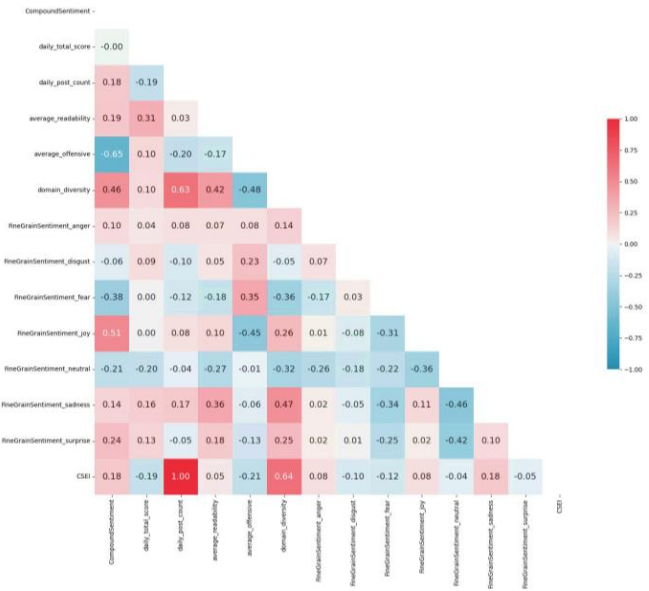

Figure 1. Correlation Matrix to show the results of correlation analysis of CSEI with each of its constituent features

Figure 2 shows an analysis of cumulative changes in CSEI with respect to major COVID-19-related events. During the initial phases of the pandemic, the CSEI shows a swift rise, corresponding with events such as the WHO's announcement of COVID-19 as a pandemic and the onset of the first wave of worldwide lockdowns. This significant upward trend reflects a time of increased public worry, apprehension, and discussions as individuals sought information, expressed their concerns, and reflected on the possible effects of COVID-19 on everyday life. Every subsequent event - be it the relaxation of restrictions, spikes in case numbers, or advancements in vaccine development - led to swift changes in CSEI, reflecting the impact each event had on the global sentiment and engagement, which can be measured and interpreted via CSEI. For example, the approval of the Pfizer vaccine, followed by a marked increase in the CSEI, shows how positive developments related to COVID-19 associated with positive social media discourse can be interpreted using CSEI. Similarly, the number of deaths due to COVID-19 surpassing one million on a global scale is also followed by a marked increase in the CSEI. This shows that negative developments related to COVID-19 associated with negative social media discourse can also be interpreted using CSEI.

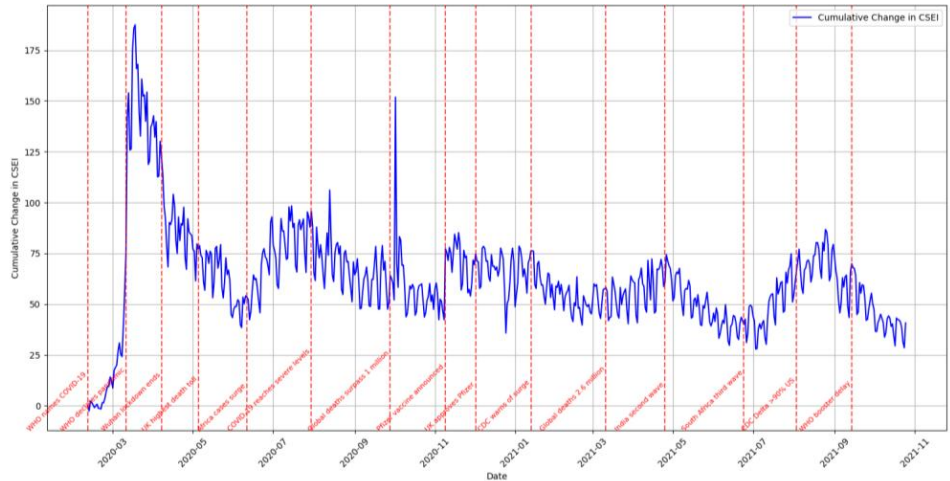

Figure 2. Cumulative Change in CSEI with respect to major COVID-19-related events

Figure 3, which highlights peaks and valleys in the rolling mean of the CSEI, offers a finer-grained view of public sentiment's responsiveness to events. Peaks represent periods of increased sentiment and engagement, often following major events related to COVID-19. For instance, the initial peaks observed after WHO declared COVID-19 a pandemic are some of the most pronounced, reflecting the widespread urgency of the situation. Similar peaks appear around critical points like the emergence of new variants, lockdowns in various countries, and significant vaccination milestones, demonstrating that public sentiment and engagement significantly increased during those events. Valleys in the CSEI rolling mean, on the other hand, represent periods when public sentiment and engagement diminished, often as the impact of previous events waned. For example, once the public had adjusted to certain restrictions or the initial shock of an event had subsided, the CSEI tended to dip, indicating a return to a baseline level of engagement and sentiment. Notably, valleys are often observed following peaks that correspond to major events, suggesting a cyclical pattern where intense public reaction is followed by periods of relative calm or normalization. By analyzing cumulative changes in tandem with peaks and valleys, it becomes evident that certain events had lasting impacts that were retained within the cumulative CSEI, while others induced short-lived reactions that dissipated quickly.

The results of Pearson's correlation analysis to determine the correlation between changes in CSEI and major COVID-19 events revealed that there was a statistically significant relationship between the two (p=0.0428). The average change in CSEI on event vs non-event days was also calculated, and the results of the same are shown in Figure 4.

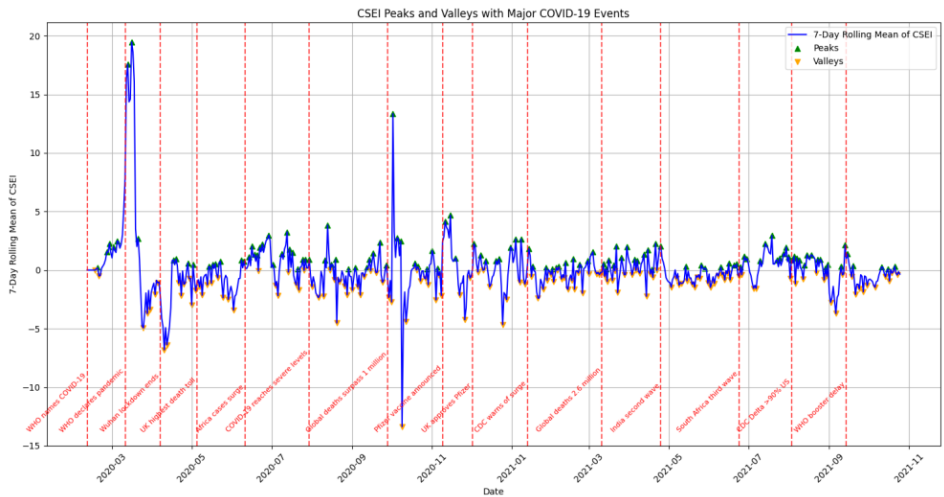

Figure 3. Evaluating CSEI Peaks and Valleys with respect to major COVID-19-related events

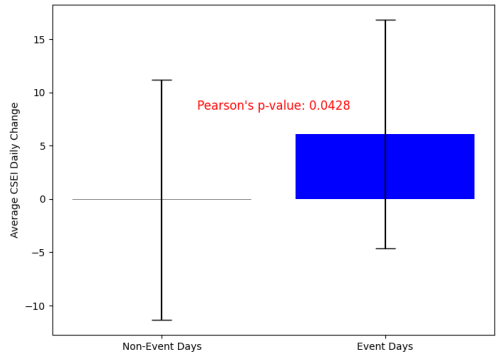

Figure 4. Average CSEI daily change on event vs. non-event days

Figure 5 highlights the effect each of the fine-grain sentiment classes - fear, surprise, joy, sadness, anger, disgust, and neutral had towards CSEI during the entire data range of the dataset, i.e., February 11, 2020, to October 25, 2021.

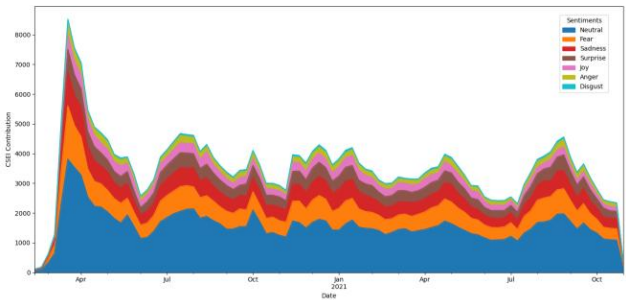

Figure 5. An analysis of the contribution of the fine fear, surprise, joy, sadness, anger, disgust, and neutral toward CSEI

As can be seen from Figures 2 and 3, peaks in CSEI align with high-intensity events, reflecting the emotional complexity of public discourse throughout various phases of the pandemic. This feature of CSEI - the capacity to mirror intricate changes in emotional tone and engagement dynamics while preserving a stable framework - allows it to provide a thorough insight into public sentiment and engagement patterns. These findings highlight the efficacy of the CSEI in accurately reflecting subtle shifts in public sentiment and engagement on social media, especially in relation to key COVID-19 events. The fixed-weight formula of the CSEI integrates multiple components - sentiment scores, engagement metrics, and detailed emotional markers - each playing a specific role in shaping the overall index. Although the weights stay constant, the actual values of these components such as fear, joy, sadness, anger, and so on - fluctuate dynamically based on the particular context and content of social media posts. The variation in component values allows the CSEI to effectively capture emotional shifts linked to major events while maintaining the integrity of the formula's structure. For example, when lockdowns were announced or COVID-19 cases increased, there was often an uptick in negative emotions such as fear and sadness [90-92]. The fixed weights in the CSEI formula indicate that an increase in these values leads to a corresponding change in the overall CSEI, emphasizing a greater level of community anxiety and distress during those times. On the other hand, during events such as vaccine rollouts, elements linked to positive emotions, like joy and a decrease in anger [93-95], showed increased values, indicating a shared feeling of relief and hopefulness. The ability to respond variably to nuanced sentiments is crucial for distinguishing the wide range of emotional reactions that arise from different events, demonstrating the CSEI's sensitivity in understanding the patterns of sentiment and engagement of the public on a global scale. This adaptable nature of the CSEI formula enables it to serve as a stable measure while still capturing the evolving public sentiment and engagement patterns. In summary, CSEI offers a robust approach for tracking and interpreting COVID-19-related public sentiment and engagement variations on social media platforms, such as Reddit.

This work has a couple of limitations. First, the methodology for the development of CSEI relied on textual data and did not take into account any images or videos that may have been included in these posts. Second, the formula for CSEI is presented based on the data available in this dataset. As COVID-19 continues to impact public health on a global scale, it is possible that major developments or events related to COVID-19 in the near future (for instance, new variants, new forms of treatment, etc.) could affect the public sentiment and engagement on social media platforms. So, if new data is collected after such major developments and the methodology described in this paper is used on the new data, the formula for CSEI computed based on such new data may vary as compared to the formula presented in this paper.

## V. CONCLUSION

The work of this paper focused on the development and evaluation of the Community Sentiment and Engagement Index (CSEI) as a metric to assess public sentiment and engagement during the COVID-19 pandemic. The CSEI combines various elements of sentiment, emotional tones, and engagement metrics, forming a comprehensive index that adapts to public reactions and engagement as they evolve over time. Utilizing a combination of compound sentiment, fine-grained sentiment, and engagement metrics like domain diversity and offensiveness, the CSEI offers a comprehensive measure that captures the intricacies of public reaction to COVID-19. The incorporation of fine-grain sentiment components improves the CSEI's ability to capture distinct emotional responses associated with various types of events, allowing for a more granular examination of how specific emotions and engagement patterns contribute to the global sentiment related to COVID-19.

This study also demonstrates how public opinion, as well as engagement patterns, evolved in response to important COVID-19 events by analyzing the change in the CSEI and examining its peaks and valleys. The findings show that major events related to COVID-19, such as WHO announcements, worldwide case surges, and vaccine rollouts, had considerable effects on public sentiment and engagement, as evidenced by sharp peaks in CSEI during those times. The cumulative analysis of CSEI presented in this paper highlights how some events left long-lasting impressions on the public mood and engagement patterns, while the peak and valley analysis of CSEI demonstrates the cyclical nature of short-term global reactions to new information and policy changes related to COVID-19. The CSEI is expected to serve as an essential resource for policymakers and health officials aiming to analyze and interpret public sentiment and engagement dynamics on social media platforms, such as Reddit, for immediate and responsive decision-making. Future work would involve extending the framework of CSEI to capture additional sentiment dimensions such as empathy, hope, despair, and trust and updating the formula accordingly.